\documentclass[aps,prd,floatfix,notitlepage,superscriptaddress,preprintnumbers,nofootinbib,10pt]{revtex4-2}

\usepackage{graphicx} 

\usepackage[T1]{fontenc} 
\usepackage{comment}
\usepackage{appendix}

\usepackage[dvipsnames]{xcolor}
\usepackage{feynmp-auto}
\usepackage{mathtools,slashed}
\usepackage{tikz}
\usepackage{amsmath}
\usepackage{amssymb}
\usepackage{cancel}
\usepackage{soul} 

\newcommand\hc{\text{h.c.}}

\begin{document}
\title{Upper limit on dark matter mass in the inert doublet model}

\author{Wararat Treesukrat}
\email{w.treesukrat@gmail.com}
\affiliation{Theoretical High-Energy Physics and Astrophysics Research Unit, Department of Physics, Srinakharinwirot University, 114 Sukhumvit 23 Rd., Wattana, Bangkok
10110, Thailand}

\author{Kem Pumsa-ard}
\email{kem@g.swu.ac.th}
\affiliation{Theoretical High-Energy Physics and Astrophysics Research Unit, Department of Physics, Srinakharinwirot University, 114 Sukhumvit 23 Rd., Wattana, Bangkok
10110, Thailand}

\author{Nopmanee Supanam}
\email{Nopmanee@g.swu.ac.th}
\affiliation{Theoretical High-Energy Physics and Astrophysics Research Unit, Department of Physics, Srinakharinwirot University, 114 Sukhumvit 23 Rd., Wattana, Bangkok
10110, Thailand}

\author{Patipan Uttayarat}
\email{patipan@g.swu.ac.th}
\affiliation{Theoretical High-Energy Physics and Astrophysics Research Unit, Department of Physics, Srinakharinwirot University, 114 Sukhumvit 23 Rd., Wattana, Bangkok
10110, Thailand}

\begin{abstract}
We study the upper limit on dark matter mass in the context of the inert doublet model. We derive an analytic expression for the upper bound as a function of the mass squared differences between dark matter and other new particles. We find that the upper limit varies between 20 and 80 TeV depending on the mass squared splitting. 
\end{abstract}

\maketitle
\flushbottom
\section{Introduction}
The discovery of the 125 GeV Higgs boson at the Large Hadron Collider ~\cite{ATLAS:2012yve,CMS:2012qbp} completes the Standard Model (SM) of particle physics. Subsequent measurements of the Higgs boson properties are consistent with the SM expectations~\cite{CMS:2022dwd,ATLAS:2022vkf}. Nevertheless, there is still room for possible deviations from the SM predictions of the Higgs boson properties. Such deviations would be a clear indication of new physics.   

One possible new physics is dark matter (DM) whose existence has been strongly hinted at by astrophysical and cosmological observations~\cite{ParticleDataGroup:2024cfk}. A recent analysis of the cosmic microwave background data indicates that the abundance of DM in the Universe is~\cite{Planck:2018vyg}
\begin{equation}
    \Omega_{DM}h^2 = 0.1200 \pm 0.0012,
\end{equation}
where $h$ is the reduced Hubble constant. It is commonly assumed that DM is a new type of particle which interacts weakly with the SM sector. Moreover, this particle must be stable on the cosmological timescale and is nonrelativistic; see Refs.~\cite{Feng:2010gw,Arcadi:2017kky,Bertone:2018krk,Arbey:2021gdg,Cirelli:2024ssz,Bozorgnia:2024pwk,Balazs:2024uyj} for recent reviews of DM. 

Even though the SM does not contain any particles with correct properties for DM, many beyond the Standard Model (BSM) scenarios can provide a DM candidate. One of the simplest and most studied BSM scenarios is the inert doublet model (IDM) where the Higgs sector of the SM is extended by an additional electroweak doublet with a stabilizing discrete $Z_2$ symmetry~\cite{Deshpande:1977rw,LopezHonorez:2006gr,Arhrib:2013ela}. The lightest neutral component of the new doublet serves as a DM candidate. In this scenario, DM interacts with the SM sector via its interaction with the Higgs boson and weak gauge interactions. As a result, DM phenomenology only depends on four free parameters: the coupling of DM to the Higgs boson and masses of the three new particles.

It has been noted in the literature that in the IDM, the correct DM abundance can be achieved if the DM mass is either lower than the $W$ mass, or heavier than 500 GeV~\cite{Arcadi:2019lka,Treesukrat:2019ahh}. The low mass region is tightly constrained by the invisible decay width of the Higgs boson measurements~\cite{ATLAS:2022yvh,CMS:2022qva}, direct DM detection experiments~\cite{LZ:2022lsv,XENON:2023cxc}, and indirect DM detection experiments~\cite{MAGIC:2016xys}. The high mass region, on the other hand, is more difficult to probe because most direct and indirect detection experiments lose sensitivity for a DM mass above the TeV scale. However, the next generation indirect detection experiment could probe DM mass up to the 100 TeV region~\cite{CTAConsortium:2017dvg}. Thus, it is imperative to determine how heavy DM can be in the IDM.

In this work, we derive analytically an upper limit on DM mass as a function of model parameters. We take into account theoretical constraints on the model parameter space, in particular, the vacuum stability and unitarity constraints. These constraints play an important role in establishing an upper bound on DM mass.

The paper is organized as follows. We briefly review the IDM in Sec.~\ref{sec:model}. In Sec.~\ref{sec:abundance}, we determine DM abundance as a function of model parameters. We then use the DM abundance to determine the upper limit on DM mass in Sec.~\ref{sec:upperbound}. Finally, we conclude in Sec.~\ref{sec:conclusion}. 
\section{The model}
\label{sec:model}
In this section, we give a brief review of the IDM. The IDM is an extension of the SM with an additional electroweak scalar doublet, $\Phi$, with hypercharge 1/2. The discrete $Z_2$ symmetry is imposed on the model under which $\Phi$ is odd while the rest of the fields are even. This $Z_2$ symmetry ensures that $\Phi$ does not participate in electroweak symmetry breaking, does not couple to the SM fermion, and the lightest neutral component of $\Phi$ is stable and can serve as a dark matter candidate. 

The scalar sector of the model is described by the potential
\begin{equation}
\begin{aligned}
	V(H,\Phi) &= -\mu _1^2H^\dagger H + \mu _2^2\Phi^\dagger\Phi + \frac{\lambda _1}{2}  (H^\dagger H)^2 + \frac{\lambda _2}{2}  (\Phi^\dagger\Phi)^2 \\
	&\qquad + \lambda _3 H^\dagger H\Phi^\dagger\Phi + \lambda_4 H^\dagger\Phi \Phi^\dagger H + \frac{\lambda _5}{2} \left((\Phi^\dagger H)^2+\hc\right),
\end{aligned}
\label{eq:potential}\end{equation}
where $H$ is the SM Higgs doublet.
The quartic coupling $\lambda_5$, in principle, could be complex. However, its phase can be absorbed into the phase of $\Phi$, rendering it unphysical. As a result, we will take all the quartic coupling real in this work. The two doublets, in unitary gauge, can be expanded as
\begin{equation}
	H = \frac{1}{\sqrt{2}}\begin{pmatrix}0\\ v+h \end{pmatrix},\qquad
	\Phi = \begin{pmatrix}\phi^+\\ \frac{\phi+iA}{\sqrt{2}} \end{pmatrix},
\end{equation}
where $v=246$ GeV is the vacuum expectation value (vev), $h$ is the 125 GeV Higgs boson, $\phi^+$ is a singly charged scalar, $\phi$ and $A$ are the \textit{CP} even and \textit{CP} odd neutral scalars, respectively. 

After electroweak symmetry breaking, the fields $h$, $\phi$, $A$, and $\phi^+$ gain masses
\begin{align}
    m_h^2 &= \lambda_1v^2,\\
	m_{\phi^+}^2 &= \mu_2^2 + \frac{\lambda_3}{2}v^2,\\
    m_\phi^2 &= \mu_2^2 + \lambda_Lv^2,\\
	m_A^2 &= \mu_2^2 + \lambda_Rv^2,
\end{align}
where $\lambda_{L,R} = (\lambda_3+\lambda_4\pm\lambda_5)/2$ is the coupling of the 125 GeV Higgs boson to the $\phi\phi$ and $AA$ pairs.
The lighter of the $\phi$ and $A$ can be a dark matter candidate, $\chi$, provided it is lighter than the singly charged scalar. Note the mass parameter $\mu_2^2$ makes it possible for the additional scalars $\phi$, $A$, and $\phi^+$ to be much heavier than the electroweak scale. Five of the seven real parameters in the scalar potential can be re-expressed in terms of physical masses and the electroweak vev, $v$. The remaining two parameters can be taken to be $\lambda_2$ and $\lambda_L/\lambda_R$. 

The quartic couplings $\lambda_i$ are constrained by perturbativity, vacuum stability, and unitarity of the partial wave amplitude. In our analysis, we take $|\lambda_i|<4\pi$ for perturbativity. Vacuum stability implies a lower bound on the couplings~\cite{Ivanov:2006yq}
\begin{align}
    \lambda_1,\,\lambda_2 &>  0, \\
    \lambda_3 &> -\sqrt{\lambda_1\lambda_2},\\
    \lambda_3 + \lambda_4 - |\lambda_5| &> -\sqrt{\lambda_1\lambda_2},
    \label{eq:vacuumstability}
\end{align}
while partial wave unitarity places an upper bound~\cite{Ginzburg:2003fe}:
\begin{align}
    3(\lambda_1+\lambda_2) + \sqrt{9(\lambda_1-\lambda_2)^2+(2\lambda_3+\lambda_4)^2} &< 16\pi,\\
    \lambda_1+\lambda_2 + \sqrt{(\lambda_1-\lambda_2)^2+4\lambda_4^2} &< 16\pi,\\
    \lambda_1+\lambda_2 + \sqrt{(\lambda_1-\lambda_2)^2+4\lambda_5^2} &< 16\pi,\\
    |\lambda_3 + 2\lambda_4 \pm 3\lambda_5| &< 8\pi,\\
    |\lambda_3\pm\lambda_4| &< 8\pi,\\
    |\lambda_3\pm\lambda_5| &< 8\pi.
\end{align}

An important consequence of perturbativity constraints is that mass squared differences $m_A^2-m_\phi^2$ and $m_{\phi^+}^2-m_\phi^2$ are of the order $v^2$. Hence, in the scenario where DM mass is much heavier than the electroweak scale, the mass of $\phi$, $A$ and $\phi^+$ are nearly degenerate. This observation will be important when determining the thermal relic density of DM in the next section. 

\section{Dark matter abundance}
\label{sec:abundance}
In the inert doublet model, the lighter of the two neutral scalars, $\phi$ and $A$, is the DM candidate. Its present day abundance is determined by its annihilation and freeze-out in the early Universe. For the heavy DM scenario, the mass of $\phi$, $A$, and $\phi^+$ is nearly degenerate. As a result, the three species freeze out at approximately the same temperature. Thanks to the discrete $Z_2$ symmetry, the heavier particles, after freeze-out, will subsequently decay into DM. Thus, the total density of DM is the sum of the density of $\phi$, $A$, and $\phi^+$. Hence, one needs to consider self-annihilation processes of $\phi$, $A$, and $\phi^+$, and coannihilation processes~\cite{Griest:1990kh} among them in determining the DM relic density. 

\subsection{Annihilation cross sections}
\begin{figure}[h]
\centering
\includegraphics[width=1\textwidth]{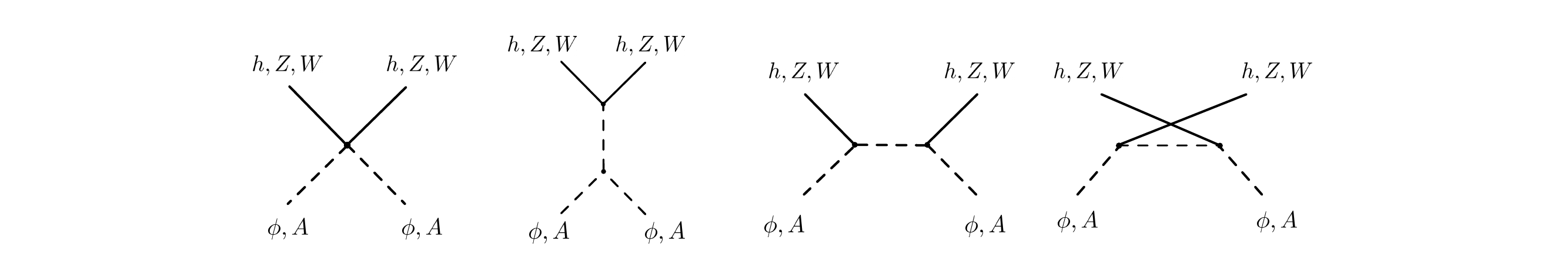}
\caption{Feynman diagrams for $\phi\phi$ and $AA$ annihilation processes.}
\label{fig:phiphixsec}
\end{figure}
 
We first work out the self-annihilation cross sections for $\phi$, $A$, and $\phi^+$ into SM particles. The full expressions for the cross sections are lengthy and not illuminating. However, since the mass of DM, $m_\chi$, is much heavier than the electroweak scale, one can expand the cross sections in terms of a small ratio, $v^2/m_\chi^2$.

The $\phi\phi$ and $AA$ annihilation processes proceed through Feynman diagrams shown in Fig.~\ref{fig:phiphixsec}. In the case of $\phi\phi$ annihilation, the cross sections are given by
\begin{align}
    \sigma v_{\text{rel}}(\phi\phi\to WW) &=  \frac{1}{8\pi m_\phi^2}\left[\left(\lambda_L + \frac{m_{\phi^+}^2-m_\phi^2}{v^2}\right)^2 + \frac{2m_W^4}{v^4}\right],\\
    \sigma v_{\text{rel}}(\phi\phi\to ZZ) &=  \frac{1}{16\pi m_\phi^2}\left[\left(\lambda_L + \frac{m_A^2-m_\phi^2}{v^2}\right)^2 + \frac{2m_Z^4}{v^4}\right],\\
    \sigma v_{\text{rel}}(\phi\phi\to hh) &= \frac{\lambda_L^2}{16\pi m_\phi^2},
\end{align}
where $m_W$ and $m_Z$ are the masses of the $W$ and $Z$ gauge bosons respectively.
Note that in the above cross sections, as well as the cross sections below, we have dropped terms which are suppressed by $v^2/m_\chi^2$ and/or are $p$ wave suppressed. This includes the $\phi\phi$ annihilation to SM fermion pairs. The $AA$ annihilation cross sections can be obtained from the $\phi\phi$ cross sections via the replacement $m_\phi \leftrightarrow m_A$ and $\lambda_L\to\lambda_R$. 

\begin{figure}[ht!]
\centering
\includegraphics[width=1\textwidth]{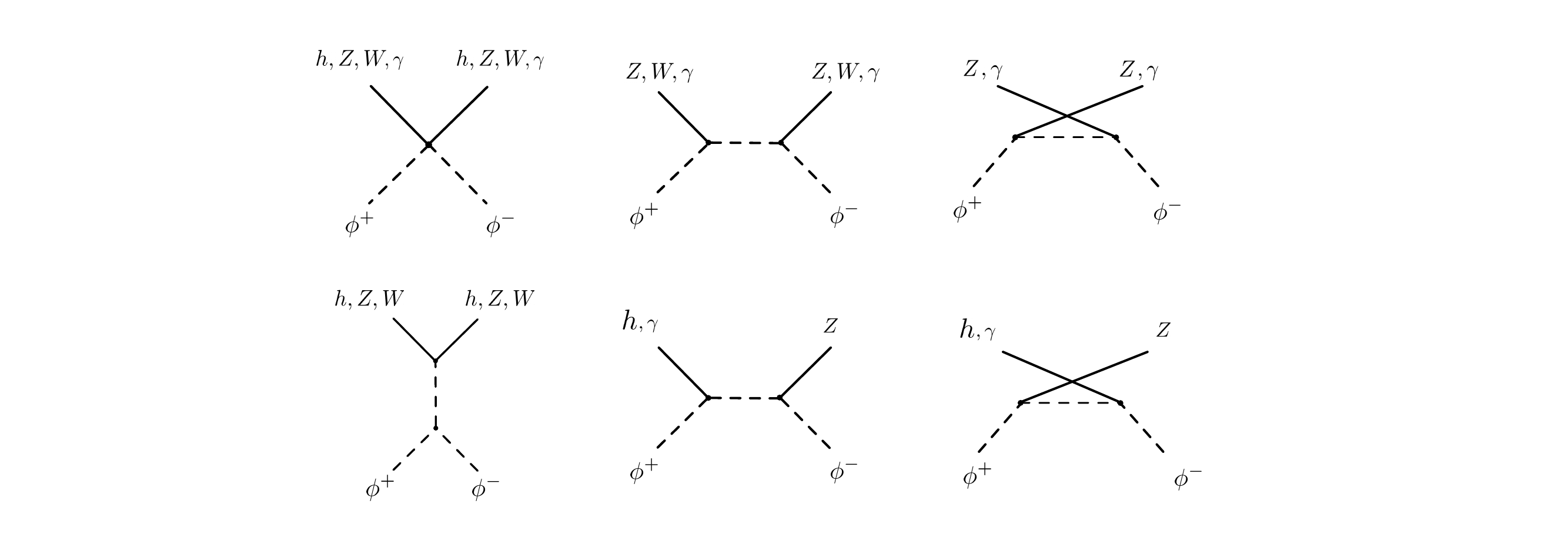}
\caption{Feynman diagrams for $\phi^+\phi^-$ annihilation processes.}
\label{fig:phiplusphiminus}
\end{figure}
The $\phi^+\phi^-$ annihilation processes are mediated by Feynman diagrams shown in Fig.~\ref{fig:phiplusphiminus}, and are given by
\begin{align}
    \sigma v_{\text{rel}}(\phi^+\phi^-\to WW) &= \frac{1}{8\pi m_{\phi^+}^2} \left[\left(\lambda_3 + \frac{m_A^2-m_{\phi^+}^2}{2v^2}+\frac{m_\phi^2-m_{\phi^+}^2}{2v^2}\right)^2 + \frac{2m_W^4}{v^4}\right],\\
    \sigma v_{\text{rel}}(\phi^+\phi^-\to ZZ) &= \frac{1}{16\pi m_{\phi^+}^2} \left(\frac{\lambda_3^2}{4} + 2c_{2w}^4\frac{m_Z^4}{v^4}\right),\\
    \sigma v_{\text{rel}}(\phi^+\phi^-\to hh) &= \frac{1}{16\pi m_{\phi^+}^2} \left(\lambda_L + \frac{m_{\phi^+}^2-m_\phi^2}{v^2}\right)^2,\\
    \sigma v_{\text{rel}}(\phi^+\phi^-\to hZ) &= \frac{c_{2w}^2\lambda_3^2}{8\pi m_{\phi^+}^2},\\
    \sigma v_{\text{rel}}(\phi^+\phi^-\to \gamma\gamma) 
    &=\frac{s_{2w}^4}{8\pi m_{\phi^+}^2}\frac{m_Z^4}{v^4},\\
    \sigma v_{\text{rel}}(\phi^+\phi^-\to \gamma Z) &= \frac{c_{2w}^2s_{2w}^2}{4 \pi m_{\phi^+}^2}\frac{m_Z^4}{v^4} ,
\end{align}
where $s_{2w}$ ($c_{2w}$) is the sine (cosine) of twice the Weinberg angle. As in the $\phi\phi$ and $AA$ annihilation cases, the $\phi^+\phi^-$ annihilation into SM fermion pairs is dropped because they are both $v^2/m_\chi^2$ and $p$ wave suppressed.

\begin{figure}[ht!]
\centering
\includegraphics[width=1\textwidth]{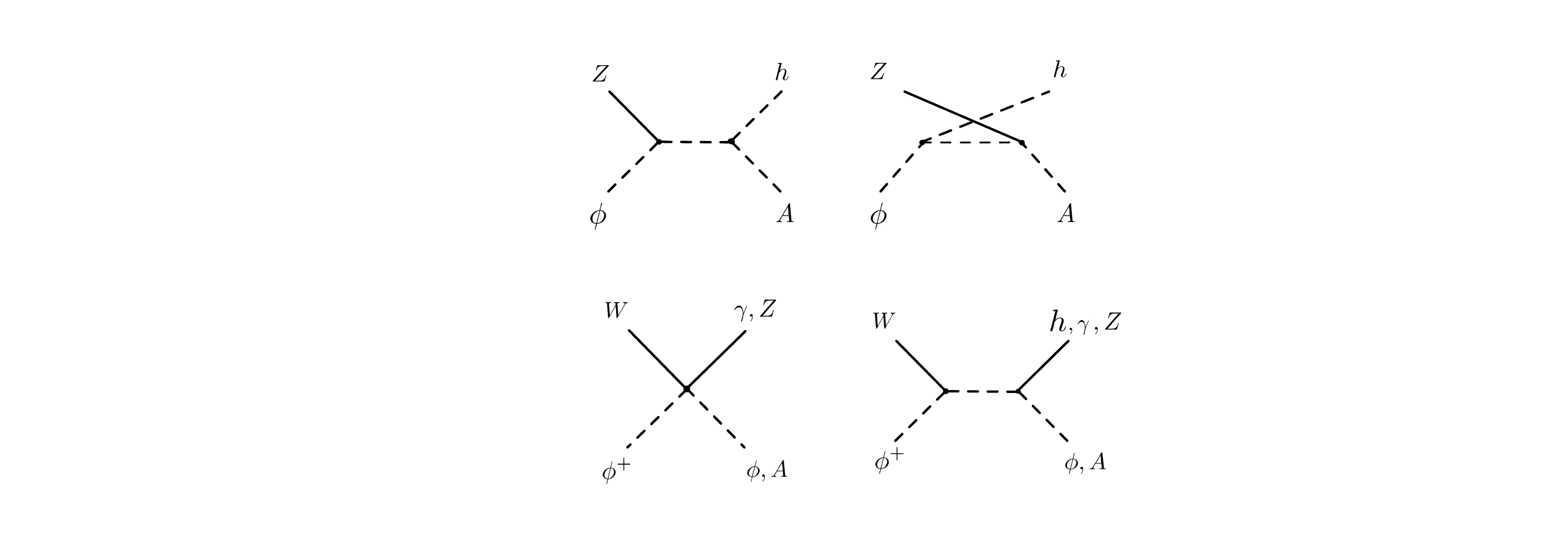}
\caption{Feynman diagrams for the coannihilation processes.}
\label{fig:coannihilation}
\end{figure}

In addition to the self-annihilation processes, $\phi$, $A$, and $\phi^+$ can annihilate with each other into SM particles. These processes are commonly referred to as coannihilation. The coannihilation cross sections proceed via Feynman diagrams shown in Fig.~\ref{fig:coannihilation}. For $\phi^+\phi$ coannihilation, the cross-sections are given by
\begin{align}
    \sigma v_{\text{rel}}(\phi^+ \phi \to W h) & = \frac{(\lambda_3- 2\lambda_L)^2}{8\pi \left(m_{\phi^+}+m_\phi\right)^2},\\
    \sigma v_{\text{rel}}(\phi^+ \phi \to WZ) &= \frac{1}{8\pi(m_{\phi^+}+m_\phi)^2}\left(\frac{\left[(1-c_{2w})(m_{\phi^+}^2-m_\phi^2) - 2(m_{\phi^+}^2-m_A^2)\right]^2}{4v^4} + \frac{2(1-c_{2w})^2m_W^2m_Z^2}{v^4}\right),\\
    \sigma v_{\text{rel}}(\phi^+ \phi \to W \gamma) &=  \frac{s_{2w}^2}{4\pi (m_{\phi^+}+m_\phi)^2}\frac{m_W^2m_Z^2}{v^4}.
\end{align}
The $\phi^+A$ coannihilation cross sections can be obtained from $\phi^+\phi$ cross sections by the replacement $m_\phi\leftrightarrow m_A$ and $\lambda_L\to\lambda_R$. Finally, the $\phi A$ coannihilation cross section is given by
\begin{equation}
   \sigma v_{\text{rel}}(\phi A \to Z h) = \frac{\lambda_5^2}{2\pi \left(m_\phi+m_A\right)^2}.
\end{equation}

\subsection{Dark matter relic density}
The evolution of $\phi$, $A$, and $\phi^+$ densities, owing to the presence of coannihilation, is governed by a set of coupled Boltzmann equations
\begin{equation}
    \frac{d n_i}{dt} = -3Hn_i - \sum_j\langle\sigma_{ij}v_{\text{rel}}\rangle\left(n_in_j - n_i^{eq}n_j^{eq}\right),
\end{equation} 
where $i = \{\phi,A,\phi^+\}$ is the specie index, $H$ is the Hubble constant, $\langle\sigma_{ij}v_{\text{rel}}\rangle$ is the thermal average annihilation cross section between specie $i$ and $j$ into SM particles, and $n_i^{(eq)}$ is the (equilibrium) number density of specie $i$. The density of DM, $n_\chi$, is the sum of the density of each scalar, $n_{\chi} = \sum n_i$. Thus, the density of DM is determined by
\begin{equation}
    \frac{d n_\chi}{dt} = -3Hn_\chi - \sum_{i,j}\langle\sigma_{ij}v_{\text{rel}}\rangle\left(n_in_j - n_i^{eq}n_j^{eq}\right).
\end{equation}

In principle, the coupled Boltzmann equations can be solved numerically. However, we will follow the approximations introduced in Ref.~\cite{Griest:1990kh} to obtain a simple analytic solution. First, the ratio of the specie-$i$ density to the DM density, to a good approximation, maintains its equilibrium value during freeze-out. Second, since $\phi$, $A$, and $\phi^+$ are nonrelativistic, their equilibrium densities can be approximated by 
\begin{equation}
    n^{eq} \sim g (mT)^{3/2}e^{-m/T},
\end{equation}
where $g$ is the number of degrees of freedom and $T$ is the temperature. It is convenient to define the dimensionless variables $\delta_i\equiv (m_i-m_\chi)/m_\chi$ and $x\equiv m_\chi/T$. Now using $n_i/n_\chi = n_i^{eq}/n_\chi^{eq}$ and the nonrelativistic approximation for $n^{eq}$, the above Boltzmann equation can be written as
\begin{equation}
    \frac{d n_\chi}{dt} = -3Hn_\chi - \langle\sigma_\text{eff}v_{\text{rel}}\rangle\left[n_\chi^2 - (n_\chi^{eq})^2\right],
    \label{eq:Boltzmann}
\end{equation}
where 
\begin{equation}
    \sigma_\text{eff} = \sum_{ij}\sigma_{ij}\frac{g_ig_j}{g_\text{eff}^2}(1+\delta_i)^{3/2}(1+\delta_j)^{3/2}e^{-x(\delta_i+\delta_j)},
\end{equation}
with 
\begin{equation}
    g_\text{eff} = \sum_k g_k(1+\delta_k)^{3/2}e^{-x\delta_k}.
\end{equation}
Equation ~\eqref{eq:Boltzmann} is in the form of the standard 1-specie Boltzmann equation. Its solution can be easily obtained by the freeze-out approximation, where the density $n_\chi$ tracks the equilibrium density $n_\chi^{eq}$ until the freeze-out temperature is reached. After that, the density $n_\chi$ becomes approximately frozen while $n_\chi^{eq}$ drops exponentially. The freeze-out temperature $T_f$, or equivalently $x_f = m_\chi/T_f$, is given by
\begin{equation}
    x_f = \log \frac{0.038\,g_\text{eff}\,m_{Pl}\,m_\chi\langle\sigma_\text{eff} v_{\text{rel}}\rangle}{g_*^{1/2}x_f^{1/2}},
    \label{eq:freeze-out}
\end{equation}
where $m_{Pl} = 1.22\times10^{19}$ GeV is the Planck mass and $g_*$ is the effective relativistic degrees of freedom at the time of freeze-out. Then the present day energy density of DM is given by
\begin{equation}
    \Omega_\chi h^2 \simeq \frac{1.07\times10^{9} \text{ GeV}^{-1}}{J g_*^{1/2}m_{Pl}},
\end{equation}
where $J$ is the depletion of $DM$ after freeze-out and is given by
\begin{equation}
    J = \int_{x_f}^\infty \frac{\langle \sigma_\text{eff}v_{\text{rel}} \rangle}{x^2}dx.
\end{equation}
In our scenario, $\langle \sigma_\text{eff} v\rangle$ does not depend on $x$. Thus, the final DM energy density can be written as
\begin{equation}
    \Omega_\chi h^2 = \frac{1.07\times10^{9}\text{ GeV}^{-1}}{g_*^{1/2}m_{Pl}}\frac{x_f}{\langle\sigma_\text{eff}v_{\text{rel}}\rangle}.
    \label{eq:DMdensity}
\end{equation}

For a given set of model parameters, one can use Eqs.~\eqref{eq:freeze-out} and \eqref{eq:DMdensity} to determine the corresponding DM relic density. Alternatively, one can use Eqs.~\eqref{eq:freeze-out} and \eqref{eq:DMdensity} to determine the nominal value of $\langle\sigma_\text{eff}v_{\text{rel}}\rangle$, the so-called thermal relic cross section, that reproduces the observed DM density as a function of $m_\chi$. Adopting this latter approach, we have found the thermal relic value for $\langle\sigma_\text{eff}v_{\text{rel}}\rangle$ to be around $2.6\times10^{-26}$ cm$^3$/s for DM mass 500 GeV $\lesssim m_\chi\lesssim10^5$ GeV. Our value is in the middle between the precise calculation value of $2.2\times10^{-26}$ cm$^3$/s presented in Ref.~\cite{Steigman:2012} and the nominal value of $3.0\times10^{-26}$ cm$^3$/s often quoted in the literature. For our numerical analysis below, we will take the thermal relic cross section to be in the range $(2.2-3.0)\times10^{-26}$ cm$^3$/s to account for the uncertainties in the thermal relic cross section.

\section{Upper limit on dark matter mass}
\label{sec:upperbound}
The five free parameters in the model can be chosen to be the DM mass $m_\chi$, two mass squared differences between the new Higgs bosons and DM, the quartic coupling $\lambda_L/\lambda_R$, and the quartic coupling $\lambda_2$. Of the five free parameters, $\lambda_2$ is the only parameter which does not have a direct impact on DM phenomenology. In this section, we will determine the maximum possible value of $m_\chi$ as a function of the two mass squared differences.

For heavy $m_\chi$, the dimensionless variable $\delta_i$ in the previous section is of order $v^2/m_\chi^2$. Since we are working to lowest order in $v^2/m_\chi^2$, $\delta_i$ can be dropped. As a result, we can approximate $g_\text{eff} \simeq \sum_kg_k = 4$ and $\sigma_\text{eff} \simeq \sum_{ij}\sigma_{ij}g_ig_j/16$. These approximations, together with the thermal relic value of the cross section, allow us to determine the upper bound on $m_\chi$ analytically. 

We first consider the scenario where $\phi$ is the DM candidate. In this case, we take the free parameters to be $m_\chi=m_\phi$, $\lambda_L$ and the two mass squared differences 
\begin{align}
    \Delta^2_0 &\equiv m_A^2 - m_\phi^2 = -\lambda_5v^2,\\
    \Delta^2_+ &\equiv m_{\phi^+}^2 - m_\phi^2 = -\frac{\lambda_4+\lambda_5}{2}v^2.
\end{align}
Note that since $m_\phi$ is the lightest, both $\lambda_5$ and $\lambda_4+\lambda_5$ must be negative. The  coupling $\lambda_R$ can be written as $\lambda_R = \lambda_L + \Delta_0^2/v^2$. In terms of the free parameters, $\sigma_\text{eff}$ is given by
\begin{equation}
    \sigma_\text{eff} = \frac{1}{16\pi m_{\chi}^2}\left[(3+2c_{2w}^2)\lambda_L^2 +\frac{5\Delta_0^2 + (14+16c_{2w}^2)\Delta_+^2}{4v^2}\lambda_L + R\right]
    \equiv \frac{f(\lambda_L,\Delta_0^2,\Delta_+^2)}{16\pi m_\chi^2},
    \label{eq:sigmaeff}
\end{equation}
where
\begin{equation}
\begin{aligned}
    R v^4 &= (45-2c_{2w}+c_{2w}^2)\frac{\Delta_0^4}{64} + 
    (57-2c_{2w}+65c_{2w}^2)\frac{\Delta_+^4}{32} - 
    (1+c_{2w})^2\frac{\Delta_0^2\Delta_+^2}{32}\\ 
    &\quad+\frac{3}{2}m_W^4 + \frac{3}{4}m_Z^4 + \frac{1-c_{2w}}{2}m_W^2m_Z^2.
\end{aligned}
\end{equation}
Since $\sigma_\text{eff}$ must reproduce the thermal relic cross section, the maximum value of $m_\chi$ is attained when the function $f(\lambda_L,\Delta^2_0,\Delta^2_+)$ is maximized subject to the vacuum stability and unitarity constraints discussed in Sec.~\ref{sec:model}. Here, we list again the relevant constraints in terms of free parameters $\lambda_2$, $\lambda_L$, $\Delta_0^2$, and $\Delta_+^2$. The vacuum stability constraints are 
\begin{align}
    \lambda_2 &> 0,\label{eq:vac1}\\
    \lambda_L &> -\frac{m_h}{2v}\sqrt{\lambda_2}\label{eq:vac2},
\end{align}
while the unitarity constraints read as~\footnote{Other vacuum stability and unitarity constraints listed in Sec.~\ref{sec:model} lead to weaker bounds on $\lambda_L$.}
\begin{align}
    3\left(\frac{m_h^2}{v^2}+\lambda_2\right) + \sqrt{\left(3\left(\frac{m_h^2}{v^2}-\lambda_2\right)\right)^2 + \left(4\lambda_L + \frac{\Delta_0^2}{v^2} + \frac{2\Delta_+^2}{v^2}\right)^2} & \leq 16\pi,\label{eq:unitary1}\\
    \left|2\lambda_L +\frac{5\Delta_0^2-2\Delta_+^2 }{v^2}\right| &< 8 \pi,\label{eq:unitary2}\\
    \left|2\lambda_L +\frac{-\Delta_0^2 + 4\Delta_+^2 }{v^2}\right| &< 8\pi\label{eq:unitary3}.
\end{align}
Note that the two vacuum stability constraints, together with the first unitarity constraint, imply that $0 < \lambda_2 < 8\pi/3$.

\begin{figure}[ht!]
\centering
\includegraphics[width=0.5\textwidth]{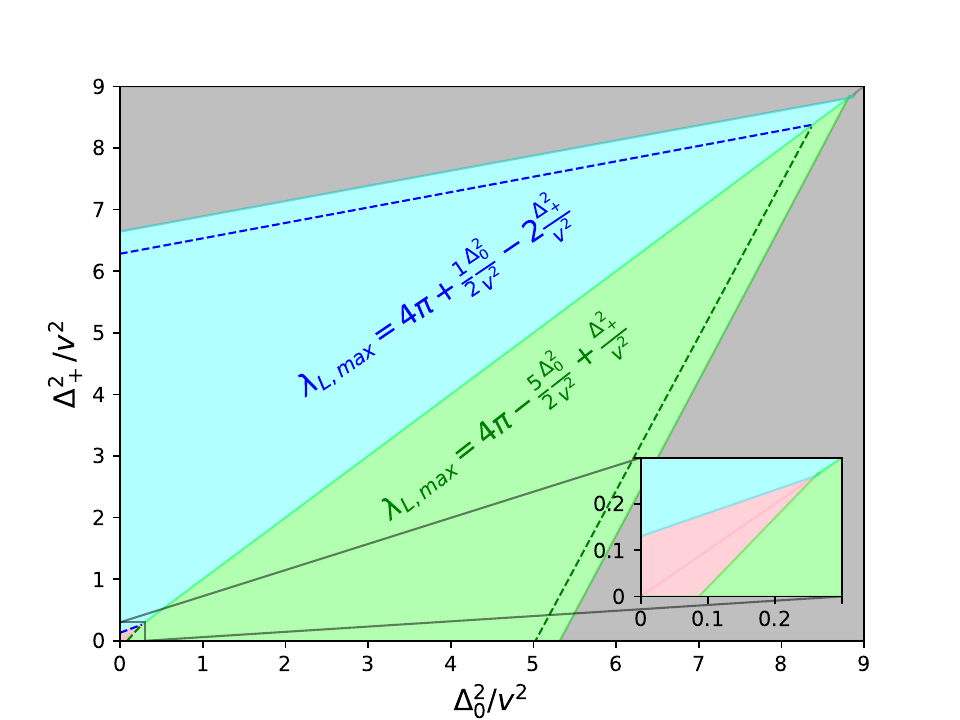}
\caption{The region on $\Delta_0^2-\Delta_+^2$ plane consistent with vacuum stability and unitarity constraints. The gray shaded area is excluded.}
\label{fig:bounds}
\end{figure}

\begin{figure}[ht!]
\centering
\includegraphics[width=0.5\textwidth]{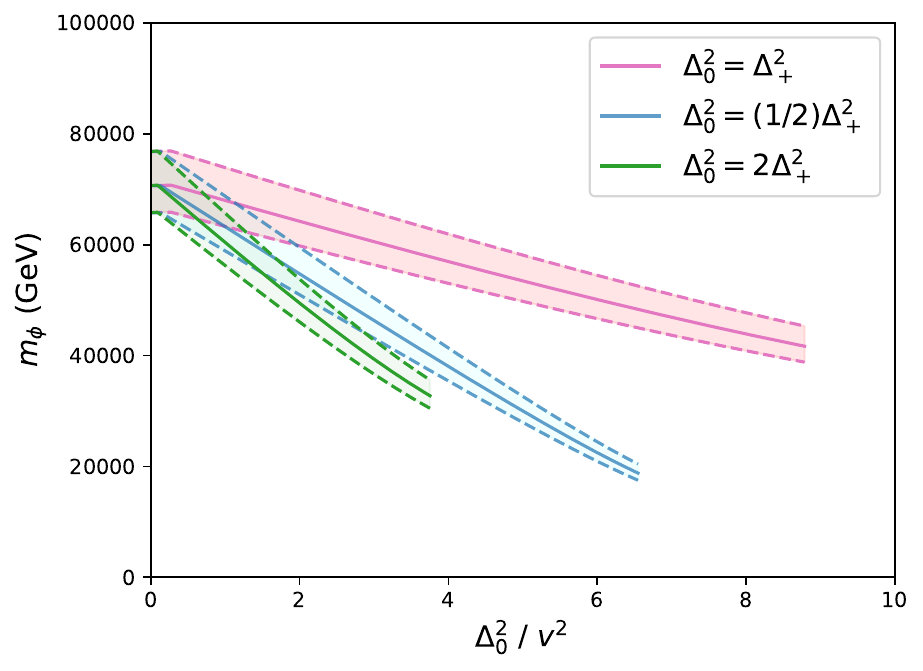}
\caption{Maximum value of $m_\chi$ for benchmark scenarios $\Delta_0^2=\Delta_+^2$ (red), $\Delta_0^2=\Delta_+^2/2$ (blue), and $\Delta_0^2=2\Delta_+^2$ (green). Solid lines represent the thermal relic cross section 2.6$\times10^{-26}$ cm$^3$/s. Color bands represent the spread due to the assumed thermal relic value between $(2.2-3.0)\times10^{-26}$ cm$^3$/s. }
\label{fig:phiDM}
\end{figure}

To maximize the function $f$, we observe that at any point on the $\Delta_0^2-\Delta_+^2$ plane, $f$ is a convex function of $\lambda_L$ with a minimum at 
\begin{equation}
    \hat\lambda_L = -\frac{5\Delta_0^2 + (14+16c_{2w}^2)\Delta_+^2}{8(3+2c_{2w}^2)v^2},
\end{equation} 
which is negative definite. Meanwhile, the region on the $\Delta_0^2-\Delta_+^2$ plane compatible with the vacuum stability and unitarity constraints are shown in Fig.~\ref{fig:bounds}. First, consider the subregion bounded by the green and blue dashed lines. On this subregion, one can always make $\lambda_L > 0$ by taking $\lambda_2$ vanishing. Thus, on this region, the function $f$ is maximized when $\lambda_L$ is at the maximum possible value allowed by unitarity constraints, $\lambda_{L,max}$. Next, consider the region bounded between the green and blue dashed lines, and the gray shaded region.  On this region, $\lambda_L$ is negative with a lower bound of
\begin{equation}
    \lambda_{L,min} = -\frac{m_h}{2v}\sqrt{\frac{8\pi}{3}} \simeq -0.74. 
\end{equation}
One can verify that $\hat\lambda_L < \lambda_{L,min}$ everywhere in this region. Hence, the function $f$ is again maximized at $\lambda_L = \lambda_{L,max}$. For the most part of the allowed region, the green and blue shaded regions of Fig.~\ref{fig:bounds}, $\lambda_{L,max}$ takes a simple form:
\begin{equation}
    \lambda_{L,max} = \left\{
    \begin{aligned}
        &4\pi - \frac{5}{2}\frac{\Delta_0^2}{v^2} + \frac{\Delta_+^2}{v^2},\\
        &4\pi + \frac{1}{2}\frac{\Delta_0^2}{v^2} - 2\frac{\Delta_+^2}{v^2},
    \end{aligned}
    \begin{aligned}
        \phantom{\frac12}&\text{ for}\phantom{\frac12} \Delta_0^2 \ge \Delta_+^2,\\
        \phantom{\frac12}&\text{ for}\phantom{\frac12} \Delta_0^2 < \Delta_+^2.
    \end{aligned}
    \right.
\end{equation}
For the part close to the origin (the pink shaded region), $\lambda_{L,max}$ is given by
\begin{equation}
    \lambda_{L,max} = 4\pi\sqrt{1-\frac{3}{8\pi}\frac{m_h^2}{v^2}} - \frac{1}{4}\frac{\Delta_0^2}{v^2} - \frac{1}{2}\frac{\Delta_+^2}{v^2}.
\end{equation}
Finally, plugging $\lambda_{L,max}$ into $\sigma_\text{eff}$, one can derive an analytic expression for the maximum $m_\chi$ as a function of $\Delta^2_0$ and $\Delta^2_+$. Figure~\ref{fig:phiDM} shows the upper limit on $m_\chi$ for three benchmark scenarios: $\Delta_0^2=\Delta_+^2$, $\Delta_0^2=\Delta_+^2/2$, and $\Delta_0^2=2\Delta_+^2$.  

We now consider the case where $A$ is the DM candidate. In this case, it is more convenient to take as free parameters $m_\chi=m_A$, $\lambda_R$ and the mass squared differences
\begin{align}
    \tilde{\Delta}_0^2 &\equiv m_\phi^2-m_A^2 = \lambda_5v^2,\\
    \tilde{\Delta}_+^2 &\equiv m_{\phi^+}^2-m_A^2 = \frac{\lambda_5-\lambda_4}{2}v^2,
\end{align}
where $\lambda_5$ is positive and $\lambda_4<\lambda_5$. The coupling $\lambda_L$ is related to $\lambda_R$ by $\lambda_L=\lambda_R + \tilde{\Delta}_0^2/v^2$. In terms of these parameters, the $\sigma_\text{eff}$ and the bounds from vacuum stability and unitarity take exactly the same form as in the previous case, with the trivial replacements $\lambda_L \to \lambda_R$, $\Delta_0^2 \to \tilde{\Delta}_0^2$, and $\Delta_+^2\to\tilde{\Delta}_+^2$. This results in the same upper limit for the DM mass as in the $\phi$ case. 

\section{Conclusion and discussion}
\label{sec:conclusion}
In this work, we have systematically analyzed the upper limit on the DM mass in the context of the IDM. We derived an analytic expression for the bound on the DM mass as a function of mass squared splittings. In the case where the neutral \textit{CP}-even $\phi$ is the DM candidate, the mass squared splittings are taken to be $\Delta_0^2 = m_A^2-m_\phi^2$ and $\Delta_+^2 = m_{\phi^+}^2-m_\phi^2$. We find that DM can be as heavy as 80 TeV for vanishing mass squared splittings. However, as $\Delta_0^2$ and/or $\Delta_+^2$ increases, the upper limit decreases.  Figure~\ref{fig:phiDM} shows three benchmark scenarios for $\Delta_0^2=\Delta_+^2$, $\Delta_0^2=\Delta_+^2/2$ and $\Delta_0^2=2\Delta_+^2$. The case where the neutral \textit{CP}-odd $A$ is the DM candidate also results in the same upper limit on DM mass as a function of mass squared differences $\tilde{\Delta}_0^2 = m_\phi^2-m_A^2$ and $\tilde{\Delta}_+^2 = m_{\phi^+}^2-m_A^2$.

In our calculation, we have repeatedly dropped contributions which are suppressed by $v^2/m_\chi^2$. This approximation allows us to obtain a compact expression for the DM annihilation cross section, and a simple solution to the coupled Boltzmann equations. The upper limits on DM mass obtained from our analysis are typically in tens TeV region. This makes our approximation well justified.

At first glance, it seems there should be no upper bound DM mass because the IDM admits the decoupling limit in which new particles can be made arbitrarily heavy. However, the DM annihilation cross section is inversely proportional to its mass squared. Thus, to reproduce the thermal relic cross section, $\langle\sigma v_{\text{rel}}\rangle \sim 2.6\times10^{-26}$ cm$^3/$s, the DM mass cannot grow without bound.

The phenomenology of the IDM with a compressed spectrum has been studied in Ref.~\cite{Tsai:2019eqi}. However, their analysis focuses on the low DM mass region, $m_{DM}\le100$ GeV, and thus does not overlap with our work. On the other hand, Refs.~\cite{Duangchan:2022jqn,Justino:2024etz} have studied the IDM in the heavy DM mass regime in the context of an indirect detection experiment. Using the micrOMEGAs package~\cite{Belanger:2014vza,Belanger:2018ccd,Alguero:2022inz}, they have obtained the upper bounds on DM mass for various benchmark scenarios which are consistent with our upper limits.


Finally, we note that such heavy DM can be probed by the next-generation gamma-ray telescopes. In particular, the upcoming Cherenkov Telescope Array (CTA) can probe the DM mass up to 100 TeV~\cite{CTAConsortium:2017dvg}. CTA sensitivity for the IDM has been studied in Refs.~\cite{Duangchan:2022jqn} and~\cite{Justino:2024etz} using dwarf spheroidal galaxies and Galactic Center as targets respectively.

\begin{acknowledgments}
K. P., N. S., and P. U. acknowledge support from the NSRF via the Program Management Unit for Human Resources \& Institutional Development, Research, and Innovation , Grant No. B39G670016. W.T. acknowledges support from the National Research Council of Thailand (NRCT): NRCT5-RGJ63017-153. P.U. also thanks the High-Energy Physics Research Unit, Chulalongkorn University for hospitality while part of this work was being completed.
\end{acknowledgments}

\bibliographystyle{unsrt}
\bibliography{ref}
\end{document}